\documentstyle[12pt,aaspp4]{article}

\def\arcmin{\ifmmode{^\prime}
    \else{$^\prime$}\fi}
\def\arcsec{\ifmmode{^{\prime\prime}}
    \else{$^{\prime\prime}$}\fi}
\def\msolar{\ifmmode{M_\odot}
    \else{$M_\odot$}\fi}
\def\h{\ifmmode{{\rm H}_2}
    \else{H$_2$}\fi}
\def\deg{\ifmmode{^\circ}
    \else{$^\circ$}\fi}

\def\slantfrac#1#2{\hbox{$\,^#1\!/_#2$}}

\def\onequarter{\slantfrac{1}{4}}

\let\tableline=\hline

\slugcomment{Accepted for publication in the Astrophysical Journal Letters}
\lefthead{Thornley}
\righthead{Spiral Structure in Flocculent Galaxies}

\begin{document}

\title{Uncovering Spiral Structure in Flocculent Galaxies}

\author{Michele D. Thornley\altaffilmark{1,2}}
\affil{Department of Astronomy, University of Maryland, College Park,
MD 20742}

\altaffiltext{1}{Visiting Astronomer, Kitt Peak National Observatory,
National Optical Astronomy Observatories, operated by the Association
of Universities for Research in Astronomy, Inc., under contract with
the National Science Foundation.}
\altaffiltext{2}{michele@astro.umd.edu}

\begin{abstract}
\noindent 

We present K\arcmin~ (2.1$\mu$m) observations of four nearby
flocculent spirals, which clearly show low-level spiral structure and
suggest that kiloparsec-scale spiral structure is more prevalent in
flocculent spirals than previously supposed.  In particular, the
prototypical flocculent spiral NGC~5055 is shown to have regular,
two-arm spiral structure to a radius of 4~kpc in the near infrared,
with an arm-interarm contrast of 1.3.  The spiral structure in all
four galaxies is weaker than that in grand design galaxies.  Taken in
unbarred galaxies with no large, nearby companions, these data are
consistent with the modal theory of spiral density waves, which
maintains that density waves are intrinsic to the disk. As an
alternative, mechanisms for driving spiral structure with
non-axisymmetric perturbers are also discussed.  These observations
highlight the importance of near infrared imaging for exploring the
range of physical environments in which large-scale dynamical
processes, such as density waves, are important.

\end{abstract}


\section{Introduction}

Flocculent galaxies are most noted for what they lack: large-scale,
continuous spiral structure, such as that identified in the grand
design spirals M51 and M81.  This fundamental difference, generally
credited to the absence of spiral density waves, makes flocculent
galaxies an important class of objects in the study of galactic
structure. Consequently, much theoretical and observational work has
focused on the physical conditions which affect structure formation in
flocculent galaxies.  If flocculent spirals have no active spiral
density waves in their stellar disks (e.g., \markcite{EnE84}Elmegreen
\& Elmegreen 1984,\markcite{EnE85} 1985), star formation is left to
occur stochastically (e.g., \markcite{gs78}Gerola \& Seiden 1978),
thereby forming only localized structures.  This assertion has been
supported by numerical simulations, which show that model galaxies
lacking density wave modes form only small- and intermediate-scale
structures (e.g., \markcite{EnT93}Elmegreen \& Thomasson 1993,
\markcite{rob93}Roberts 1993).

This theoretical picture is largely based on observations at optical
wavelengths; however, studies of the distribution of dust in galaxies
suggest that extinction significantly affects the observed structure
at wavelengths as long as 0.8$\mu$m (I band, see
e.g. \markcite{rixr93}Rix \& Rieke 1993).  In addition, optical bands
are dominated by emission from recent star formation, which does not
necessarily trace the underlying stellar potential.  With a
significant contribution from old stars (\markcite{rix93}Rix 1993) and
as much as an order of magnitude less extinction than at optical
bands, near infrared (NIR) imaging of galaxies can provide significant
new constraints on models of spiral structure.  Indeed, variations in
spiral structure between optical and near infrared images of grand
design spirals have recently been observed (e.g., Rix \& Rieke 1993,
Block et al. 1994).  However, weakly structured disk galaxies are just
beginning to be examined in detail at NIR wavelengths (e.g., M33,
Regan \& Vogel 1994, hereafter RV94).

We present K\arcmin~ CCD imaging of four nearby flocculent galaxies:
NGC~2403, NGC~3521, NGC~4414, NGC~5055.  These images show spiral
structure on kiloparsec scales, enlarging the sample of galaxies in
which regular, spiral structure is detected.  Future papers will
better define the physical significance of the spiral structure
reported here, through comparison with the distribution of gas
(molecular + atomic) and recent, massive star formation.  In \S2 the
K\arcmin~ observations and data reduction are described; in \S3 the
method for highlighting nonaxisymmetric structures by subtracting
model bulge and disk components is explained; in \S4 the observed
spiral structure is described; and in \S5 possible origins of
spiral structure in flocculent galaxies are discussed.

\section{Observations and Data Reduction Techniques}

Infrared images in the K\arcmin~(2.1~$\mu$m) broadband filter
(\markcite{wc92}Wainscoat \& Cowie 1992) were obtained using the
Cryogenic Optical Bench (COB) on the 1.3m telescope of the Kitt Peak
National Observatory during 1-5 March 1994.  COB uses a 256x256 InSb
detector, which at the time of the observations had 0.94\arcsec~pixels
on the 1.3m and a 4\arcmin~field of view.  The effective resolution
was 2.0\arcsec. The K\arcmin~data were taken as 20-second exposures
and coadded before readout for an exposure time of 120 seconds.
Exposures were taken in dithered pairs, followed by a pair of dithered
sky frames taken 15-20 arcminutes away from the galaxy center. The
galaxy center was then moved successively around the array, in order
to produce an 8\arcmin~x~8\arcmin~ or (in the case of NGC~2403)
12\arcmin~x~12\arcmin~mosaiced image.  The total exposure time per
galaxy was approximately 50 minutes.

Infrared standard stars (\markcite{elias}Elias et al. 1982;
\markcite{ch92}Casali \& Hawarden 1992) were observed over a range of
airmasses to calibrate the infrared images and derive extinction
coefficients.  A single flat field was created each night by median
filtering all dark-subtracted sky exposures. Individual sky frames
were made for each pair of object frames by median filtering the sky
frames taken immediately before and after the object frames.  For each
object frame, the composite sky frame was subtracted, a flat field
correction was applied, and a correction for extinction was made.  The
individual dithered object frames were registered using stars found in
the overlap regions from frame to frame, and a $\chi^2$ minimization
technique (\markcite{regg95}Regan and Gruendl 1995) was used to create
the final mosaic.  A correction for the zero point of each mosaic
image was made by observing a series of overlapping fields which
extended sufficiently far from the galaxy to be free of emission.
Astrometry was performed using field stars from the Hubble Guide Star
Catalog.

\section{Bulge and Disk Decomposition} \label{NIRmod}

The resulting K\arcmin~mosaic images (Figure 1) show the predominantly
axisymmetric light distribution expected in flocculent galaxies.
However, at low levels non-axisymmetric structure is detectable,
extending to 0.2-0.4~R$_{25}$.  To better characterize the structures
seen in K\arcmin~ emission, modeling and subtraction of axisymmetric
components were undertaken.  The process is similar to the analysis
completed for M33 by \markcite{regv94}RV94: a radial profile
constructed by measuring the brightness in concentric annuli is fit
simultaneously with two model components, a spherical or flattened de
Vaucouleurs r$^{\onequarter}$ bulge and an exponential disk. The best
fit models are then subtracted from the mosaic images.  To minimize
the contribution from a bright nuclear component (three of the sample
galaxies are potential LINERS, \markcite{heck80} Heckman 1980,
\markcite{keel83}Keel 1983), the inner 4 pixels (3.8\arcsec) were
excluded from the fit.  To minimize the contribution of spiral arms,
the radial profile was constructed from the 25th percentile of the
brightness distribution in each annulus.

This simple method of disk-bulge decomposition is used only to remove
a smooth, axisymmetric luminosity component; a more detailed model
should be used in order to accurately characterize the related mass
distribution.  The flattened r$^{\onequarter}$ bulge model inherently
assumes an oblate bulge with the same apparent axial ratio as the
disk, the value of which was determined from the ellipticity of
isophotes within the inner 50\arcsec.  Models with spherical and
flattened r$^{\onequarter}$ bulges cannot be distinguished by $\chi^2$
using only a radial profile, but subtraction of a spherical model from
K\arcmin~images of all sample galaxies except NGC~2403 showed large
negative residuals along the minor axis, suggesting a non-spherical
bulge. Finally, while radial profiles cannot distinguish between
spiral arms and other deviations of the light distribution from that
of the chosen models, it is not possible to create non-axisymmetric
structure in the residual images by subtracting purely axisymmetric
models.  After subtraction of axisymmetric models, the residuals in
all galaxies are less than 0.2 magnitudes near the ``arm'' structures,
indicating that the subtraction of model components is sufficient to
characterize low-level spiral structure, which is the focus of this
Letter.

The inclination, bulge effective radius, and disk scale length for each
galaxy are shown in Table~\ref{fits}.  To indicate the quality
of the fits, the radial profiles of NGC~4414 and NGC~5055 are compared
with the model fits in Figure 2.  The radial profile of NGC~2403 shows
little emission in excess of an exponential disk, and thus the disk
was fit first and the bulge fit to the remaining structure in the
profile.  In NGC~4414, the small angular size of the galaxy allowed
few radial points across the bulge, leading to larger uncertainties in
the bulge fit.  In the radial profile of NGC~5055, the effects of the
approximation of a simple, two-component model are seen: the bulge and
disk components have comparable brightness contributions at most radii
covered by the profile (Figure 2b).  The radial profile of NGC~5055
has a contribution from a strong ring enhancement at
r$\sim$40\arcsec~(see \S4.1), which encourages a fit of a stronger
central component.  Disk and bulge models were subtracted from the
original K\arcmin~images to produce the residual images shown in
Figure 3.  The relatively uncertain bulge contribution in NGC~2403 can
be seen in the slight oversubtraction of the inner disk (Figure 3a).
Small bar-like residuals in the central $\sim$15\arcsec~ are due to
uncertainties in the axial ratio of the bulge for these simple models.
There is no evidence for strong, kiloparsec-scale bars in the NIR.

\section{Spiral structure in flocculent galaxies} \label{spiral}

The distribution of the residual emission presented in Figure 3
suggests the presence of two components in the NIR spiral structure.
The distribution of small-scale, bright features correlates well with
the distributions of HII regions detected in H$\alpha$ imaging
(Thornley et al. 1996a, in prep.), suggesting a contribution from a
younger population, such as K supergiants (e.g. \markcite{rixr93}Rix
\& Rieke 1993, \markcite{regv94}RV94).  However, the presence of a
broader, smooth component of emission in the residual maps affirms the
higher sensitivity of NIR emission to the smoother distribution of the
old stellar population. In the following sections, we describe the
structure seen in the residual images.

\subsection{\bf Individual Galaxies}

{\bf {\it NGC~5055.}} The Sbc galaxy NGC~5055 (d=7.2~Mpc) is one of the
prototypes for an Arm Class 3 flocculent galaxy
(\markcite{EnE87}Elmegreen \& Elmegreen 1987), but it shows clearly a
symmetric, two-arm spiral structure in the NIR (Figure 3d).  Each arm
extends over 150\deg~ in azimuth to a radius of $\sim$2.0\arcmin~
(4~kpc, 0.3~R$_{25}$) before decreasing in brightness.  A ring of
excess K\arcmin~ emission at a radius of 40\arcsec~ is coincident with
a slight enhancement of HII regions seen in H$\alpha$ images (Thornley
et al. 1996, in prep.).  The bulge axial ratio indicated by inner
isophotes is 0.67, while the axial ratio of the disk at larger radii
is 0.57, suggesting that the bulge is triaxial.

{\bf {\it NGC 2403.}}  NGC~2403 is a nearby (3.2~Mpc) Scd galaxy with
similar global properties to the Local Group spiral M33.  The NIR
residual structure in NGC~2403 is relatively uncertain due to its low
surface brightness and minimal bulge, but a two-arm spiral structure
is marginally detected (see Table 2), extending over as much as
180\deg~ in azimuth to a radius of $\sim$2~kpc (0.2~R$_{25}$).
The stronger, outer ridges of the spiral structure are also consistent
with enhanced star formation at the radius at which the gas surface
density in the disk becomes supercritical for star formation (Thornley
\& Wilson 1995).

{\bf {\it NGC 3521.}} The residual map of the Sbc galaxy NGC~3521
(d=7.2~Mpc) shows a tightly wound two-arm pattern (Figure 3b). The
eastern arm is long and continuous, extending over 180\deg~in azimuth
to 3.5~kpc (0.3~R$_{25}$), while the western arm appears to split in
two, leaving a continuous spiral arm over $\sim$130\deg~of azimuth to
a radius of 2.4~kpc.  K\arcmin~ emission from the central region is
extremely strong, and is offset from the centroid of the disk by
$\sim$2\arcsec~ to the west.

{\bf {\it NGC 4414.}} The high gas surface density Scd galaxy NGC~4414
(\markcite{braine93}Braine et al. 1993) is 2.5 times more distant than
NGC~3521 (\markcite{pierce94}Pierce 1994), and remains the most
flocculent of the sample in K\arcmin~emission.  Comparison with
H$\alpha$ imaging suggests that much of the residual K\arcmin~emission
in the inner disk (Figure 3c) can be traced to the distribution of HII
regions in NGC~4414; for example, bright star forming regions to the
northwest and southeast of the nucleus likely contribute to an
apparent NIR ring/arm structure of radius 20\arcsec~(2~kpc).  Outer
``arm'' segments to the north and south extend to a radius of
$\sim$40\arcsec~(0.4~R$_{25}$), and are continuous over $\sim$60\deg~
in azimuth.  While these outer structures are similar in linear scale
to the arm structures seen in NGC~3521 and NGC~5055, they do not
appear to contribute to a regular two-arm spiral pattern.

\subsection{ NIR Arm-Interarm Contrasts}

To compare the stellar density enhancement along the arms with those
of grand design spirals, we have measured the arm-interarm contrast in
each galaxy, using the arms defined by the residual images.  In
NGC~5055 this was accomplished using two methods.  In the first
method, ten regions along each arm were matched with two interarm
positions at the same galactocentric radius in the original image.  In
the second method, the ten selected arm regions were compared with the
axisymmetric model (disk + bulge) at the same position.  For both
methods, the median brightness was measured in a 9~x~9 pixel region
around each position, and the arm-interarm contrast, R, is given by
the ratio of the arm to non-arm (interarm or axisymmetric model)
brightness.  The median is used to minimize the effects of localized
enhancements of K\arcmin~emission due to a younger population of K
supergiants. Due to more tightly wound spiral arms or bright foreground
stars seen in the K\arcmin~ images, the first method is only feasible
in NGC~5055.  As it is model-independent, this method is most
desirable, but by using both methods in NGC~5055 we confirm that the
results are the same to within the uncertainties (see Table 2).  The
NIR arm-interarm contrasts reported here (Table 2) range from
1.1-1.4.  Though the methods for measuring the relative
overdensity of the spiral arms are heterogeneous, galaxies which
exhibit clear grand design structure (e.g., M51, Rix 1993; M83,
Adamson et al. 1987; and NGC~7309 and IC~2627, Rix \& Zaritsky 1995)
have similarly measured NIR arm-interarm contrasts of $\sim$1.5-3.0,
suggesting that flocculent spirals have lower stellar density
enhancements than grand design spirals.

\section{The Origin of Low-Amplitude Spiral Structure} \label{discussion}

The images presented in Figure 3 demonstrate the presence of
kiloparsec scale spiral structure in nearby flocculent galaxies.
While these data do not rule out the existence of truly flocculent
galaxies, they suggest that spiral density enhancements may be present
in many more (see also NGC~1309 and NGC~1376, \markcite{rix95}Rix \&
Zaritsky 1995).  As the detected arm structure extends past the
turnover radius of the rotation curve in NGC~3521 and NGC~5055
(\markcite{CvG91}Casertano \& van Gorkom 1991, \markcite{kent87}Kent
1987), it is unlikely that the arm structures in these galaxies are
merely material arms maintained by solid-body rotation in the inner
disk (\markcite{kornor79}Kormendy \& Norman 1979, RV94); indeed, given
the short dynamical time scales in the inner disk, the regular spiral
structure detected in the inner 2~kpc of NGC~2403 also suggests a
long-term organizing influence.  However, significant spiral structure
is detected only in the inner regions of these four galaxies
(r$\le$0.4~R$_{25}$), in contrast to grand design galaxies such as M81
where the arms begin at r$\sim$0.2~R$_{25}$ and extend to large radii.

The galaxies in this sample are relatively isolated, and there is no
evidence for strong kiloparsec-scale bars; what, then, is the origin
of the low-level spiral structure in these galaxies?  The presence of
spiral structure in isolated flocculent galaxies supports the modal
theory of spiral density waves (e.g., Bertin et al. 1989a,b; Lowe et
al. 1994), which suggests that spiral structure is self-excited and
maintained by feedback between the stellar disk and the more
dissipative, self-gravitating gas disk.  Bertin et al. (1989a,b)
suggest that tightly wound spiral structure is formed relatively
easily in galaxies with low active disk mass, such as those with a
large spheroid component.  This is consistent with the observation of
the most continuous spiral structure in NGC~3521 and NGC~5055, which
have more significant bulges.  In the context of the modal theory, low
arm-interarm contrasts suggest that the conditions in the stellar and
gaseous disks are not well coupled, and therefore are unable to
support larger amplitude waves.

The low amplitude waves could also indicate that spiral structure in
flocculent galaxies is weakly driven by another component.  A slightly
triaxial bulge, if rotating, could drive low-amplitude spiral
structure in the inner disk; indeed, the suggestion that most galaxies
have nonaxisymmetric bulges (\markcite{zlo86}Zaritsky \& Lo 1986;
\markcite{bertola90}Bertola, Vietri, \& Zeilinger 1991) indicates that
this mechanism could be generally applicable to low-level spiral
structure. However, the observation of {\it rounder} inner isophotes
suggests that a weak triaxial structure in NGC~5055 would be aligned
with the minor axis, while the spiral arms appear to begin near the
major axis (Figure 3d).  A more detailed model of the bulge in
NGC~5055 is needed to analyze its effect on disk structure.  Low mass
companions may provide an alternative low-level driving mechanism;
indeed, NGC~5055 has two small potential companions (NGC~5023 and
UGC~8313, \markcite{fisht81}Fisher \& Tully 1981).  Simulations of
interactions and mergers by \markcite{bh92}Byrd \& Howard (1992) and
\markcite{mh94}Mihos \& Hernquist (1994) suggest that small companions
can induce spiral structure over the entire disk, which persists for
as long as a few billion years.  \markcite{bh92}Byrd \& Howard (1992)
further suggest that the companion tidally perturbs the outer disk,
and these perturbations in turn excite longer-lived density waves in
the inner disk.  However, neither simulation definitively constrains
the amplitude or the extent of tidally induced spiral structure.
While the ubiquity of small companions makes this mechanism attractive
in explaining the origin of low-level spiral structure in seemingly
isolated galaxies, further study is needed to ascertain whether it is
consistent with the data presented here.

\section{Summary}

In this Letter, we have demonstrated the presence of low-amplitude,
two-arm spiral structure in the inner disks of a sample of nearby
flocculent galaxies, most notably in the prototypical flocculent
galaxy NGC~5055.  The observed spiral structure may be the result of
self-excited spiral density waves, or the influence of a rotating,
slightly triaxial bulge.  Alternatively, the observed spiral structure
may be driven by tidal interaction with low-mass companions; however,
further study is needed to provide firmer support for this mechanism.
All models discussed here suggest that regular spiral structure in
NGC~2403, NGC~3521, and NGC~5055 is not particularly unique, and
low-level spiral structure in flocculent galaxies could be common.
The near infrared images presented here identify these galaxies as
weaker counterparts to grand design spirals, and detailed studies of
the ISM and star formation in future papers will help define the
properties of this sample of flocculent galaxies and gauge the
significance of density wave activity in different galactic
environments.

Thanks to Michael Regan and Robert Gruendl for assistance with
near-infrared observations. Thanks also to Christine Wilson, Lee
Mundy, and Stuart Vogel for helpful discussions during the preparation
of this manuscript, and to the anonymous referee for insightful
suggestions. This work was supported by NSF grant AST~93-14847 and a
Scholar Award from PEO International.

\clearpage

\begin{table}
\begin{center}
\caption{Axisymmetric model fits}
\label{fits}
\begin{tabular}{lccc}
Galaxy & i & R$_{e}$\tablenotemark{b} & l$_{d}$ \tablenotemark{c}\\
\tableline NGC~2403 & 60\deg & 2.5\arcmin~$\pm$0.6\arcmin~ &
1.4\arcmin$\pm$0.2\\ NGC~3521 & 57\deg\tablenotemark{a} &
0.9\arcmin$\pm$0.3\arcmin~ & 0.7\arcmin$\pm$0.1 \\ NGC~4414 &
55\deg\tablenotemark{a} & 0.1\arcmin$\pm$0.3 & 0.3\arcmin$\pm$0.05\\
NGC~5055 & 48\deg\tablenotemark{a} & 1.6\arcmin~$\pm$0.6 &
0.8\arcmin$\pm$0.1\\
\end{tabular}
\end{center}
\tablenotetext{a}{Determined from image isophotes.}
\tablenotetext{b}{Radius inside which half of the bulge light is found.}
\tablenotetext{c}{Disk scale length.}
\end{table}

\begin{table}
\begin{center}
\caption{Arm-Interarm Contrasts}
\label{arm_contrast}
\begin{tabular}{llc}
Galaxy & Arm  & R$\tablenotemark{a}$ \\
\tableline
NGC 2403 & E arm  &  1.16$\pm$0.06 \\
& W arm  &  1.24$\pm$0.07 \\
NGC~3521 &  NE arm  &  1.25$\pm$0.05 \\
& SW arm  &  1.13$\pm$0.03 \\
NGC~4414 & N ``arm''  &  1.38$\pm$0.04\\
& S ``arm''  &  1.13$\pm$0.03 \\
NGC~5055$\tablenotemark{b}$ & NE arm & 1.29$\pm$0.06\\
& SW arm & 1.33$\pm$0.06\\
\end{tabular}
\end{center}
\tablenotetext{a}{Determined from second method. See \S4.3. All
uncertainties are uncertainties in the mean value of R.} 
\tablenotetext{b}{R determined from first method: 1.28$\pm$0.04 for
the NE arm, and 1.23$\pm$0.04 for the SW arm. See \S4.3.}
\end{table}

\clearpage

\begin{figure}
\figurenum{1}
\caption{K\arcmin~images of nearby flocculent galaxies: (a)NGC~2403,
(b)NGC~3521, (c)NGC~4414, and (d)NGC~5055.  Scale bar at lower right
corresponds to distance along the major axis.}
\figurenum{2}
\caption{Plot of K\arcmin~radial surface brightness profiles for
NGC~4414 and NGC~5055. The crosses indicate the measured radial
profile, the dashed line indicates the model disk, and the dotted line
indicates the model bulge. The solid line indicates the total model
distribution.}
\figurenum{3}
\caption{Residual K\arcmin~images, after subtraction of axisymmetric
bulge and disk models from the images shown in Figure 1.  The scale
for these images is an order of magnitude smaller than in Figure
1. White crosses mark locations of foreground stars on the face of the
galaxy. (a)NGC~2403; (b)NGC~3521; (c)NGC~4414; and (d)NGC~5055. Scale
bar is shown at lower right.  See \S~3 for details.}

\end{figure}


\clearpage
\begin{references}

\reference{adam87} Adamson, A.J., Adams, D.J., \& Warwick,
R.S. 1987,\mnras, 224, 367 
\reference{bert89a} Bertin, G., Lin, C.C., Lowe, S.A. \& Thurstans,
R.P. 1989, \apj, 338, 78 
\reference{bert89a} Bertin, G., Lin, C.C.,
Lowe, S.A. \& Thurstans, R.P. 1989, \apj, 338, 104
\reference{bertola90} Bertola, F., Vietri, M., \& Zeilinger,
W.W. 1991, \apjl, 374, L13 
\reference{block94} Block, D.L., Bertin,
G., Stockton, A., Grosb$\o$l, P., Moorwood, A.F.M., \& Peletier,
R.F. 1994, \aap, 288, 365
\reference{braine93} Braine, J., Combes, F., \& van Driel, W. 1993,
\aap, 280, 451
\reference{bh92} Byrd, G.G. \& Howard, S. 1992, \aj, 103, 1089
\reference{ch92} Casali, M. \& Hawarden, T. 1992, 
The JCMT-UKIRT Newsletter, 4, 33 
\reference{CvG91} Casertano, S. \& van Gorkom, J.H. 1991, \aj, 101, 1231 
\reference{elias} Elias, J.H., Frogel, J.A., Matthews,
K., \& Neugebauer, G. 982, \aj, 87, 1029 
\reference{EnE85} Elmegreen, B.G. \& Elmegreen, D.M. 1985, \apj, 288, 438
\reference{EnT93} Elmegreen, B.G. \& Thomasson, M. 1993, \aap, 272, 37
\reference{EnE84}  Elmegreen, D.M. \& Elmegreen, B.G. 1984, \apjs, 54, 127
\reference{EnE87} Elmegreen, D.M. \& Elmegreen, B.G. 1987, \apj, 314, 3
\reference{gs78}  Gerola, H. \& Seiden, P.E. 1978, \apj, 223, 129
\reference{fisht81} Fisher, J.R. \& Tully, R.B. 1981, \apjs, 47, 139
\reference{heck80} Heckman, T.M. 1980, \aap, 87, 152
\reference{keel83} Keel, W.C. 1983, \apjs, 52, 229
\reference{kent87} Kent, S.M. 1987, \aj, 93, 816
\reference{kornor79} Kormendy, J. \& Norman, C.A. 1979, \apj, 233, 539
\reference{lowe94}Lowe, S.A., Roberts, W.W., Yang, J., Bertin, G., \& Lin,
C.C. 1994, \apj, 427, 184
\reference{mh94} Mihos, J. C. \& Hernquist, L. 1994, \apjl, 425, L13.
\reference{pierce94} Pierce, M.J. 1994, \apj, 430, 53
\reference{regg95} Regan, M.W. \& Gruendl, R.A. 1995, Astronomical Data
Analysis and Systems IV, ed. R.A. Shaw,  H.E. Payne, \& J.J.E. Haynes
(ASP Conf. Ser., 77), 335
\reference{regv94} Regan, M.W. \& Vogel, S.N. 1994, \apj, 434, 536
\reference{rix93} Rix, H.-W. 1993, \pasp, 105, 999
\reference{rixr93} Rix, H.-W. \& Rieke, M.J. 1993, \apj, 418, 123
\reference{rix95} Rix, H.-W. \& Zaritsky, D. 1995, \apj, 447, 82
\reference{rob93} Roberts, W.W. Jr. 1993, \pasp, 105, 670
\reference{tw95} Thornley, M.D. \& Wilson, C.D. 1995, \apj, 421, 458
\reference{wc92}  Wainscoat, R.J., \& Cowie, L.L. 1992, \aj, 103,332
\reference{zlo86} Zaritsky, D., \& Lo, K.-Y. 1986, \apj, 303, 66
\end{references}
\end{document}